\documentclass[12pt,preprint]{aastex}

\def\lapp{\ifmmode\stackrel{<}{_{\sim}}\else$\stackrel{<}{_{\sim}}$\fi}
\def\gapp{\ifmmode\stackrel{>}{_{\sim}}\else$\stackrel{>}{_{\sim}}$\fi}

\newcommand{\xte}{\textit{RXTE}} 
\newcommand{\tfe}{1E~1048.1--5937} 
\newcommand{\tfn}{1E~2259+586}

\newcommand{\soe}{1RXS~1708--4009} 
\newcommand{\oft}{4U~0142+61}
\newcommand{\efo}{1E~1841--045}

\newcommand{\tnin}{\mathrm{T}_{90}}

\begin{document}

\title{A Comprehensive Study of the X-ray Bursts from the Magnetar Candidate \tfn} \author{ Fotis P. Gavriil\altaffilmark{1} ,
Victoria M. Kaspi \altaffilmark{1,2,3}, Peter
M. Woods\altaffilmark{4,5}}

\altaffiltext{1}{Department of Physics, Rutherford Physics Building,
McGill University, 3600 University Street, Montreal, Quebec, H3A 2T8,
Canada}

\altaffiltext{2}{Department of Physics and Center for Space Research,
Massachusetts Institute of Technology, Cambridge, MA 02139}
\altaffiltext{3}{Canada Research Chair, Steacie Fellow, CIAR Fellow}

\altaffiltext{4}{Space Science Research Center, National Space Science
and Technology Center, Huntsville, AL 35805, USA}
\altaffiltext{5}{Universities Space Research Association}

\begin{abstract}
We present a statistical analysis of the X-ray bursts observed from the
2002 June 18 outburst of the Anomalous X-ray Pulsar (AXP) \tfn,
observed with the Proportional Counter Array (PCA) aboard the
\textit{Rossi X-ray Timing Explorer}.  We show that the properties of
these bursts are similar to those of
Soft Gamma-Repeaters (SGRs).  The similarities we find are the burst
durations follow a log-normal distribution which peaks at 99~ms, the
differential burst fluence distribution is well described by a power law
of index $-1.7$, the burst
fluences are positively correlated with the burst durations, the
distribution of waiting times is well described by a log-normal
distribution of mean 47~s, and the bursts are generally asymmetric with
faster rise than fall times.  However, we find several quantitative
differences between the AXP and SGR bursts.  Specifically, the AXP
bursts we observed exhibit a wider range of durations, the
correlation between burst fluence and duration is flatter than for
SGRs, the observed AXP bursts are on average less energetic than observed SGR bursts, and
the more energetic AXP bursts have the hardest spectra -- the opposite of
what is seen for SGRs.  We conclude that the bursts are sufficiently
similar that AXPs and SGRs can be considered united as a source class 
yet there are some interesting
differences that may help determine what physically  differentiates the 
two closely related manifestations of neutron stars.
\end{abstract}


\section{INTRODUCTION}

Soft gamma repeaters (SGRs) are an exotic class of Galactic sources
that are now commonly accepted as being magnetars -- isolated, young
neutron stars that are powered by the decay of an ultra-high magnetic
field.  The evidence for high surface fields ($\sim 10^{14} -
10^{15}$~G) comes from several independent lines of reasoning
\citep{dt92a,pac92,td95,td96a}.  These include: the high dipolar
magnetic fields implied by the spin properties of SGRs seen in
quiescence  under the assumption of magnetic dipole braking 
\citep{kds+98,ksh+99}; the requirement of a
magnetar-strength field to confine the energy released in the tails of
hyper-Eddington outbursts seen from two SGRs \citep{mgi+79,hcm+99};
the requirement of a high field to allow the decay rate necessary to
power the burst and persistent emission \citep{td96a, gr92}; 
and the magnetic suppression of the
Thomson cross-section, which allows hyper-Eddington bursts to be
observed \citep{pac92}.
%and the time scale of the initial spike of the 1979 March 5 
%flare being the stellar Alfv\'en crossing time for magnetar-strength fields.
For reviews of SGRs, see \citet{kou99b}, \citet{hur00} and \citet{tho01}.

Anomalous X-ray pulsars (AXPs), another exotic class of Galactic
neutron stars, have also been suggested to be magnetars \citep{td96a}.
This is because of their anomalously bright X-ray emission which can
be explained neither by conventional binary accretion models nor
rotation power \citep{ms95}. Also, their spin parameters, as for
SGRs, imply large magnetic fields under standard assumptions of
magnetic braking.  They also have similar, though on average softer,
X-ray spectra compared with those of SGRs in quiescence. However, unlike SGRs, 
in the $>20$~yr since the
discovery of the first AXP \citep{fg81}, none was seen to exhibit SGR-like
bursts.  For this reason, alternative models involving unconventional
accretion scenarios have been proposed to explain AXP emission
\citep{vtv95,chn00,alp01}.  See \citet{ims02} and \citet{mcis02} for 
reviews of AXPs.

The magnetar model for AXPs was recently given a boost
when SGR-like bursts were detected from two AXPs.  \cite{gkw02}
reported on the discovery of two X-ray bursts in observations
obtained in the direction of AXP \tfe.  The temporal and spectral
properties of those bursts were similar only to those seen only in SGRs.
However, the AXP could not be definitely identified as the burster.
On 2002 June 18, a major outburst was detected unambiguously from AXP \tfn,
involving over 80 bursts as well as significant spectral and timing
changes in the persistent emission \citep{kgw+03}.  Those bursts
demonstrated that AXPs are capable of exhibiting behavior observed,
until now, uniquely in SGRs, therefore implying a clear connection
between the two source classes.  Such a connection was predicted only
by the magnetar model \citep{td96a}.  However, the physical difference
between the source classes is as yet unclear; \citet{gkw02} and
\citet{kgw+03} suggest that AXPs have higher surface magnetic fields
than do SGRs, in spite of the evidence to the contrary from their
spin-down properties.

In this paper, we consider the statistical properties of the \tfn\
bursts in detail, in order to compare them quantitatively with SGR
bursts, both to confirm that they have properties sufficiently
similar that the two phenomena can definitely be unified, as well as
to look for subtle differences that may offer clues regarding the
physical distinction between the two classes.  Statistical studies of
magnetar bursts \citep[e.g.][]{gwk+99,gwk+00,gkw+01} have the potential
to yield important information regarding the burst energy injection and
radiation mechanisms.  Correlations between different burst
properties, whether temporal and spectral, can be powerful model
discriminators.  Burst statistical properties can be compared with
other physical phenomenon in order to assist in identifying their
underlying cause; for example, they have been used to argue for
important similarities between SGR bursts and earthquakes
\citep{cegy96}.

In this paper we present a comprehensive analysis of the
properties of the bursts seen in the 2002 June 18 outburst of \tfn.  
We present a study of the detailed outburst and post-outburst properties
of the persistent and pulsed emission of \tfn\ in a companion paper
\citep{wkg+03}.

\section{OBSERVATIONS AND ANALYSIS}
\label{sec:observations}
The results presented here were obtained using the Proportional Counter
Array \citep[PCA;][]{jsg+96} on board the \textit{Rossi X-ray Timing Explorer}
(\xte ). The PCA consists of an array of five collimated xenon/methane
multi-anode proportional counter units (PCUs) operating in the
2--60~keV range, with a total effective area of approximately
$\rm{6500~cm^2}$ and a field of view of $\rm{\sim 1^o}$~FWHM. We use
\xte\ to monitor all five known AXPs on a regular basis as part of a
long-term monitoring campaign \citep[see][and references therein]{gk02}. 
On 2002 June 18, during one of our
regular monitoring observations (\xte\ observation identification
70094-01-03-00) that commenced at UT 15:39:18, the AXP \tfn\ exhibited an SGR-like outburst
\citep[see Fig.~\ref{fig:light curve};][]{kgw+03}. 
The bursting behavior was detected by online \xte\ monitors during
the observation, and is clearly visible in the PCA ``Standard~1'' data.
The observation spanned three orbits and had 
total on-source integration time 10.7~ks.  Although some PCUs turned
on/off during our observation, there were exactly three PCUs
operational at all times.  In addition to the standard data
modes, data were collected in the \texttt{GoodXenonwithPropane} mode,
which records the arrival time (with 1-$\mu$s resolution) and energy 
(with 256-channel resolution) of every unrejected xenon event as well as all the
propane layer events. Processing of these data was done using software
that operated directly on the the raw telemetry data.
 Photon arrival times were adjusted to the solar
system barycenter using a source position of (J2000) RA 23$^{\rm h}$
01$^{\rm m}$ 08$^{\rm s}$.295, DEC +58$^{\circ}$ 52$'$ 44$''.45$ \citep{pkw+01} and the JPL
DE200 planetary ephemeris.
Note that following the outburst, Target of Opportunity observations of the source 
were initiated the next day and continued at different intervals over the subsequent weeks, 
however no more bursts were seen.

\subsection{The Burst Identification Algorithm}
\label{sec:searching}
To study the bursts quantitatively, we made use of the
{\texttt{GoodXenonwithPropane} data.  Time series were created
separately for each PCU using all xenon layers. Light curves of
various time bin widths ($1/1024$~s, 1/256~s, 1/64~s, 1/32~s  and 1/16~s) 
were created to allow sensitivity to bursts on a range of time scales.  The
\texttt{FTOOL}s \texttt{xtefilt} and \texttt{maketime} were used to
determine the intervals over which each PCU was off.  We further
restricted the data set by including only events in the energy range
2--20~keV.  We used this energy range, which is larger than that used
to study the quiescent pulsations \citep{gk02,wkg+03}, because of the much
harder spectra of the bursts relative to the quiescent emission.

The following procedure was performed separately for each
PCU, in order to identify bursts. 
First, for each data set, the number of counts in the $i^{\mathrm{th}}$  time bin 
was compared to a local mean $\mu_i$. The local mean was calculated over a
$\sim$28~s (four pulse periods) stretch of data centered around the time bin being
evaluated. A window of $\sim$7~s (one pulse cycle) was also 
administered so that
counts directly from, and immediately around, the point under
investigation would not contribute to the local mean.  
During the outburst there was an increase in the pulsed flux \citep{kgw+03,wkg+03},  
such that coherent pulsations were visible in our binned light curves. 
Because of this, for example, the apparent significance 
of bursts falling near a pulse
peak would be artificially enhanced.
To compensate for this effect, we first 
modelled the counts per time bin due to pulsations as:
\begin{equation}
p_i =  A(\phi_i, t_i)\left[ C e^{-t_i/\tau}\right] ,
\end{equation}
where $A(\phi,t)$ is the normalized amplitude of the pulsations as a function 
of pulse phase $\phi$ and time $t$. The parameters $C$ and  $\tau$ are 
from an exponential fit to the pulsed flux evolution.
We then calculated an adjusted local mean  in the following way:
\begin{equation}
\lambda_i = \mu_i + p_i -\sum_j p_j ,
\end{equation}
where the index $j$ spans the windowed stretch of data used to calculate the local mean.
For the number of
counts in a time bin ($n_i$) greater than the adjusted  local mean ($\lambda_i$), the
probability of those counts occurring by random chance is given by
\begin{equation}
P_i  = \frac{\lambda_i^{n_i}e^{-\lambda_i}}{n_i!},
\label{eq:burst prob pcu}
\end{equation}
As the probability $P_i$ for each PCU is independent, we calculated the
total probability ($P_{tot}$) of observing a burst simultaneously by
all operational PCUs as
\begin{equation}
P_{i,\mathrm{tot}}=\prod_{k=0}^{4}P_{i,k},
\label{eq:burst prob total}
\end{equation}
and $k$ corresponds to the PCU under consideration. If a particular PCU were inoperable we set $P_{i,k}=1$.
Events which registered a value of $P_{i,\mathrm{tot}} \le 0.01/N$, where $N$ is the total number of time bins searched,  were flagged
as bursts, and were subject to further investigation.

The significance of the number of counts in a time bin can be underestimated 
if  there are one or more bursts  in the 
interval used as the local mean. 
For this reason, once a burst was identified it 
was  removed from the light curve, and the  burst identifying 
procedure was repeated until there were no additional bursts returned.
\section{RESULTS}
\label{sec:analysis and results}

\subsection{Burst Statistics}
\label{sec:statistics}

Our burst searching algorithm returned 80 significant bursts 
from the 2002 June 18 observation. The number of bursts identified 
depended on the time resolution 
used: $26\%$, $55\%$, $76\%$, $83\%$ and $74\%$ of all identified bursts 
were flagged at 1/1024~s, 1/256~s, 1/64~s, 1/32~s  and 1/16~s  
time resolution, respectively. Most bursts 
were single-peaked and had durations
$\lapp$1~s.  A small handful ($\sim$12) were bright and had clear
fast-rise, exponential decay morphology.  Four
bursts were  multi-peaked.  A variety of burst
morphologies is shown in Figure~\ref{fig:burst profiles}. Some
bursts ($\sim$5$\%$) were approximately symmetric, 
a few ($\sim$3$\%$) fell faster than they rose
while most  fell slower than they rose (see \S\ref{sec:rise fall}).

\subsubsection{Burst Event Times and Phase}
\label{sec:time}

The time of each burst was initially defined, using binned light curves,
to be the midpoint of the bin having the most counts. 
To increase the precision of the burst time we refined this value, using the event data 
which comprised this time bin, to be the midpoint of the times of the events 
having the smallest temporal separation.  We also calculated the occurance in 
pulse phase for each burst using the time of the burst peak and the rotational
  ephemeris given by \citet{kgw+03}.
Comparing  the burst phase distribution to the pulse 
profile of \tfn\  at the time of the outburst, a correlation 
is seen (Fig.~\ref{fig:burst phases}), where most of the  bursts 
tend to occur when the pulsed intensity is high. We note that the two  bursts seen from the AXP \tfe\ \citep{gkw02} were also coincident with the pulse peak, which  strengthens the argument that \tfe\  was the source of those bursts.

\subsubsection{Burst Durations and Fluence}
\label{sec:duration}
The $\tnin$ duration is the time
between when $5\%$ and $95\%$ of the total background-subtracted burst
counts have been accumulated \citep[e.g.][]{gkw+01}.  The background count rate was determined
by averaging a hand-selected  burst-free region before and after the
burst.   This typically consisted of two intervals of 1~s before and
after the burst in question.  The integrated background-subtracted
counts were then fit to a step function plus a linear term using
least-squares fitting. The height of the step-function corresponds to
the total burst fluence $F$ (in counts) and the slope of the line
corresponds to any background counts that were improperly subtracted.

SGR $\tnin$ distributions follow a log-normal distribution, defined as
\begin{equation}
P(\tnin,\mu,\hat{\sigma}) = \frac{1}{\log\hat{\sigma}\sqrt{2\pi}}
%e^{-\frac{1}{2\log\sigma^2}\left(\log\tnin - \log\mu\right)^2},
\exp\left[ - \frac{1}{2} \left(  \frac{\log\tnin - \log\mu } {\log\hat{\sigma}}   \right)^2 \right]
\label{eq:lognormal}
\end{equation}
whose mean and standard deviation    vary with source \citep[e.g.][]{gkw+01}. 
At first we fit the measured values of $\tnin$ for the
\tfn\ bursts with this model and found it to characterize the
distribution well.  
In Equation~\ref{eq:lognormal} the parameters $\log\mu$ and $\log\hat{\sigma}$ 
correspond to the mean and standard deviation of the the $\log\tnin$ values.
The mean of the $\tnin$ values is given by $\mu$ and  the range
for one standard deviation 
corresponds to  $(\mu\hat{\sigma}^{-1}, \mu\hat{\sigma})$.
The best-fit  $\mu$ and $\hat{\sigma}$ were
determined by maximum likelihood testing.  
The latter allowed us to extract
model parameters that are independent of the arbitrarily chosen
histogram bin widths. Specifically, the best-fit parameters were those
which maximize the statistic
\begin{equation}
\mathcal{M} = \sum_{i=1}^{N}  \log P(\mathrm{T}_{90,i},\mu,\hat{\sigma}),
\end{equation}
where $N$ is the number of bursts.
 Figure~\ref{fig:T90} shows the
distribution, and best-fit log-normal model for the measured values.
We found that our $\tnin$ distribution has mean $\mu=97.9$~ms with a 
range of 18.2--527.2~ms for one standard deviation.
Note however that for low signal-to-noise bursts, $\tnin$ can be
substantially underestimated.  We describe how we corrected for this problem
and obtained slightly modified best-fit log-normal parameters in
\S\ref{sec:t90sims} below.

The fluences measured as described above were then
grouped in equispaced logarithmic bins.  The distribution of burst
fluences is displayed in Figure~\ref{fig:fluence distribution}. The
low-end fluences are underrepresented because of sensitivity drop-off.
Excluding the points having fluence \lapp 20 PCA counts, the
distribution is well modeled by a simple power law. Using least-squares
fitting we find a best-fit power-law index of $-0.7 \pm 0.1$, which 
corresponds to a differential spectrum $\mathrm{d}N/\mathrm{d}F \propto F^{-1.7 \pm 0.1}$.
From the plot, it is clear that the fluences span approximately
two orders of magnitude.  For our calibration of the fluences in
CGS units, see \S\ref{sec:burst calibration}.

\citet{gkw+01} also find a clear correlation between burst durations
and total burst fluence. In Figure~\ref{fig:fluence versus T90}, we
plot fluence versus $\tnin$.  A correlation can clearly be seen.  To
quantify it, we grouped the $\tnin$ values in equispaced logarithmic
bins and determined group-averaged fluences for each bin.
Least-squares fitting to a simple power-law model yields $F \propto
\tnin^{+0.54 \pm 0.08}$, with reduced $\chi^2 = 1.0$.

\subsubsection{Burst Peak Fluxes}
\label{sec:flux}
Burst peak fluxes were determined from the event data using the following 
algorithm.
A box-car integrator of width 62.5~ms was translated through the event data.
The procedure began and ended when the  center of the box-car was at half a box-car width
before and after the time of the burst peak (as determined in \S~\ref{sec:time}). At each box-car step
a flux measurement was made by integrating the number of events 
and dividing by the
box-car width. The burst peak flux was assigned the largest such flux measurement.
We then grouped our peak fluxes in equispaced logarithmic bins.
The distribution of peak fluxes is shown in Figure~\ref{fig:peak flux
distribution}.  

Our burst-identifying algorithm is less
sensitive to bursts of smaller peak flux.  To compensate for this
effect, we ran the following simulation. We took a hand-selected 1-ks
long burst-free region from our observed \tfn\ light curve binned with
62.5-ms resolution. We then injected a simulated 
burst having peak flux $f_p$ at a random position in the light curve.
We modelled the burst by a top-hat function of width 62.5~ms (one time bin) 
and height $f_p\times62.5$~ms.  We then ran our burst-identifying
algorithm as described in \S~\ref{sec:searching}. We repeated this procedure
for $N_i$ iterations  and determined $N_s$,  
the number of successful burst identifications for that simulated peak flux. 
We repeated the procedure for various peak fluxes and determined the
probability of detecting a burst $P = {N_s}/{N_i}$ as a function of
peak flux $f_p$.  We found that $P$ could be well modelled by the following analytic function
\begin{equation}
P(f_p) = \frac{1}{2}\left[1 + \tanh\left(\frac{f_p-f_0}{k}\right)\right],
\end{equation}
with $f_0 = 309.84~\mathrm{cts}~\mathrm{s}^{-1}$ and $k = 58.21~ \mathrm{cts}~\mathrm{s}^{-1}$.
We then used this function 
to correct our
peak flux distribution (see Fig.~\ref{fig:peak flux distribution},
boxes).  Using least-squares fitting we found that the corrected
distribution is well modelled by a simple power law with index $-1.42
\pm 0.13$.
For our calibration of these 
peak fluxes in CGS units, see \S\ref{sec:burst calibration}.

\subsubsection{Burst Rise Times and Fall Times}
\label{sec:rise fall}
Burst rise and fall times were obtained from the event data by
maximizing the likelihood of the assumed probability
distribution
\begin{equation}
P(t) = \left\{ \begin{array}{ll} A(C_p e^{(t-t_p)/t_r} + B) & t \le
t_p \\ A(C_p e^{-(t-t_p)/t_f} + B ) & t>t_p
\end{array}  \right. ,
\label{eq:model}
\end{equation}
where $B$ represents the background count rate, $C_p$ represents the
background-subtracted count rate at the time of the burst peak $t_p$,
and $t_r$ and $t_f$ represent the burst rise and fall times,
respectively. The parameter $A$ is a normalizing factor ensuring
unit probability over the interval of interest.  This model characterized 
the bursts well -- see  the left panels of Figure~\ref{fig:burst profiles} (dotted line)  for examples. 
Burst rise and
fall time distributions are displayed in Figure~\ref{fig:rise fall},
with best-fit log-normal models 
determined via maximum-likelihood testing.  For the rise time distribution, 
we find a mean of $2.43$~ms and a range of 0.51--11.51~ms for 
one standard deviation, 
with reduced $\chi^2 = 1.3$.  
For the fall time distribution,
we find mean 13.21~ms and a range  of 3.52--49.55~ms for one
standard deviation,
and a reduced $\chi^2 = 0.2$.  In order to better quantify burst
morphologies we also show the ratio of burst rise times
to fall times ($t_r/t_f$; Fig.~\ref{fig:rise fall}).  On average,
bursts rise faster than they fall, however this is not universally
true.  Again fitting a log-normal distribution, we find  mean
0.18 and a range of 0.03--1.08  for one standard deviation, with
reduced $\chi^2 = 3.7.$  The latter fit is poor because the distribution
is clearly skewed toward shorter rise times.  The asymmetry of the typical
burst can also be seen in Figure~\ref{fig:ratio tr T90}, where the
distribution of $t_r/\tnin$ is plotted.

\subsubsection{Corrected $\tnin$ Values}
\label{sec:t90sims}

\citet{gkw+01} showed that in the low signal-to-noise regime, 
the value of $\tnin$ can be underestimated.  To account
for this, a model light curve was generated for each burst, having the form of
Equation~\ref{eq:model}. Peak flux, rise time and fall time were fixed
at the values measured for that particular burst. The simulated light
curve was then integrated and the model  duration 
($\mathrm{T}_{90,m})$ was measured by the same procedure outlined in
\S\ref{sec:duration}. We then repeated the procedure with noise
added to the simulated light curve.  The noise was drawn from
a Poissonian distribution
having mean equal to the measured background rate of the burst
under investigation.  We repeated the procedure for 200
realizations of noise.  For each iteration ($i$) we measured 
the duration ($\mathrm{T}_{90,i}$). 
The simulated durations ($\mathrm{T}_{90,i}$) were normally 
distributed and the mean of this distribution 
($\mathrm{T}_{90,s}$ ) allowed  us  to calculate  a 
correction factor  
$\mathcal{FD} \equiv  1 - \mathrm{T}_{90,m}/\mathrm{T}_{90,s}$.  
The corrected $\tnin$ distribution is shown in Figure~\ref{fig:T90}. 
The best-fit mean is 99.31~ms with  a range of 14.4--683.9~ms
for one standard deviation.

\subsubsection{Burst Waiting Times}
\label{sec:waiting time}

SGR waiting times ($\Delta T$), defined as the temporal separations of
adjacent bursts, are found to follow log-normal distributions
\citep{gwk+99,gwk+00}.  We measured the waiting time for the
\tfn\ events, excluding those interrupted by Earth occultations.
Figure~\ref{fig:waiting time distribution} displays our $\Delta T$
distribution with the best-fit log-normal model as determined by
maximum likelihood testing.  The best-fit parameters are mean of
46.7~s and a range of 10.5--208.4~s for one
standard deviation, with reduced $\chi^2=0.6$. 
We find no correlation between the burst energy, 
duration and the waiting time until the
next burst, nor with the elapsed time since the previous burst.

Note however that the burst rate clearly decreased during the observation 
(see Fig~\ref{fig:light curve}). This is made clear by the bottom panel of 
Figure~\ref{fig:waiting time distribution} which shows a correlation between 
the waiting time ($\Delta T$) and the burst peak  time ($t_p$). We fit this 
correlation to a power-law model using least-squares fitting, which reveals 
that $\Delta T = {0.11}\times{{t_p}^{0.81}}$ . This correlation implies that 
the mean of  our waiting time distribution depends on the time at which 
we started  observing the outburst. 
%However, if we extrapolate the best-fit 
%power-law  model to a few seconds before the start of the observation we 
%obtain waiting  times of the order of milliseconds, which suggests  that we 
%may have caught  the outburst near its commencent. However, from our fits to 
%the timing data, we find that the rotational glitch appears to have occured 
%siginificantly before the commencement of the observations \citep{wkg+03}. 
%This suggests that the bursting may have started well 
%after the rotational glitch.

\subsection{Burst Spectroscopy}
\label{sec:burst spectra}

\subsubsection{Individual Burst Spectra}
\label{sec:ind burst spectra}

Spectra for each burst were extracted with the 256 spectral bins over
the PCA range grouped by a factor of 4 in order to increase the
signal-to-noise ratio per spectral bin. The same background intervals
selected in measuring $\tnin$ were used in the spectral analysis (see
\S\ref{sec:duration}).  In all spectral analyses, energies below 2~keV
and above 60~keV were ignored, leaving on average 33 spectral channels
for fitting.  The regrouped spectra along with their background
estimators were used as input to the X-ray spectral fitting software
package \texttt{XSPEC}\footnote{http://xspec.gsfc.nasa.gov}.  Response
matrices were created using the \texttt{FTOOL}s \texttt{xtefilt} and
\texttt{pcarsp}.
We fit the 28 most fluent bursts with a photoelectrically 
absorbed power law of index
$\Gamma$, holding only $N_H$ fixed at $0.93\times 10^{22}$~cm$^{-2}$ \citep[the value found by][]{pkw+01}.
The distribution of spectral indices is shown in Figure~\ref{fig:gamma distribution}. We find  a mean spectral index of $\Gamma=1.35$ with  standard deviation  0.43.

\subsubsection{Hardness Ratios}
\label{sec:hardness}

\citet{gkw+01} noted that SGR bursts tend to soften with increasing
burst energy.  We studied the hardness ratio/fluence relationship by
extracting spectra and creating response matrices separately for each
burst.  Hardness ratios were defined as the ratio of the counts in the
10--60~keV band to those in the 2--10~keV band as in \citet{gkw+01}.
Also following \citet{gkw+01}, we divided the bursts into equispaced
logarithmic fluence bins and calculated a weighted average hardness
ratio for each bin. Figure~\ref{fig:hardness ratio versus fluence}
shows the weighted mean hardness ratios as a function of fluence.  A
clear positive correlation is seen.  We repeated the procedure for
different definitions of hardness ratio and found similar
correlations.  We further confirmed this trend by considering the 28
most fluent bursts for which spectral indexes $\Gamma$ could be reliably and
precisely constrained.  All had $\Gamma$ well below the mean value.

\subsubsection{Absence of Spectral Lines and the Average Burst Spectrum}

Possible spectral features have been reported in a burst from the AXP
\tfe\ \citep{gkw02} and from bursts from two SGRs
\citep{si00,iss+02,isp03}.  In no spectrum of any burst for \tfn\ did
we detect a significant feature.  In order to amplify any low-level
spectral feature common to all bursts, we combined individual burst
spectra to create a grand average spectrum. We summed the burst and
background spectra described in the previous section using the
\texttt{FTOOL} \texttt{sumpha}. Response matrices were scaled and added
using the \texttt{FTOOL} \texttt{addpha}.  In order to search for
features in the residuals, we fit the combined spectrum to a simple
photoelectrically absorbed power law. The residuals showed no evidence of 
significant spectral features.

\subsubsection{Calibrating Fluence and Flux}
\label{sec:burst calibration}

Determining peak flux and total fluence distributions in CGS units
requires spectral fitting.  However most bursts were too faint to allow
spectral parameters to be determined with interesting precision.  The
problem was worse for the peak fluxes since even the brighter bursts
generally had too few counts to meaningfully constrain the spectrum.
Therefore, we devised an alternate way of converting between PCA counts
and CGS units.  We took the spectra of the 40 most luminous bursts 
extracted over their $\tnin$ duration and fit them with
photoelectrically absorbed power laws. However this time, for
consistency, we held $\Gamma$ fixed at the mean of our spectral index
distribution.  We multiplied the flux (in units of
erg~s$^{-1}$~cm$^{-2}$) in the 2--60~keV range returned by the fit by
its respective $\tnin$ duration to obtain a fluence in erg~cm$^{-2}$.
We then considered the 2--60~keV fluence in counts as determined in
\S\ref{sec:duration} as a function of the fluence in CGS units and
determined the proportionality constant between the two using
least-squares fitting.  This constant was found to be $8.226\times10^{-12}~\mathrm{erg}~\textrm{cm}^{-2}~\mathrm{cts}^{-1}$. In \S\ref{sec:hardness} we found significant spectral evolution as a function of fluence. A change 
of 1$\sigma$ in spectral index $\Gamma$ corresponds 
to a change by a factor of $\sim1.5$ in our calibration constant.
The same procedure and constant applies for the peak fluxes. 
The CGS energy scales are shown at the top of
Figures~\ref{fig:fluence distribution} and \ref{fig:peak flux
distribution}.  The fluences in the 2--60~keV band 
range from $\sim 5 \times 10^{-11}$ to
$\sim 7 \times 10^{-9}$~erg~cm$^{-2}$. 
These imply burst energies in the range $\sim 5 \times 10^{34}$ to 
$\sim 7 \times 10^{36}$~erg, assuming isotropic emission and 
a distance of 3~kpc to the source \citep{kuy02}.
The sum total of all burst fluences is $5.6\times10^{-8}$~erg~cm$^{-2}$, corresponding to 
energy $6.0\times 10^{37}$~erg~(2--60~keV).
Peak fluxes in a 61.25-ms time bin  range
from $\sim 1 \times 10^{-9}$ to
$\sim 1 \times 10^{-7}$~erg~cm$^{-2}$~s$^{-1}$, which imply peak luminosites  
in the range $\sim 1 \times 10^{36}$ to $\sim 1 \times 10^{38}$~erg~s$^{-1}$.
On shorter time scales we find 5 bursts with peak fluxes 
which are super-Eddington.  The peak fluxes in a 1/2048~s time bin for these bursts range from $\sim 2 \times 10^{38}$ to $\sim 8\times 10^{38} $~erg~s$^{-1}$.

\section{DISCUSSION}
\label{sec:discussion} 

As we describe below, many of the properties of the bursts seen from
\tfn\ during its 2002 June 18 outburst are very similar to those seen
in SGRs 1806$-$20 and 1900+14.  However, there are some quantitative
differences.  Next we compare the various measured quantities for the
AXP and SGR bursts.  Note that our comparisons focus primarily on PCA
observations of SGRs for consistency of spectral and temporal response.

The mean $\tnin$ value of 99.31~ms (see \S\ref{sec:t90sims}
and Fig.~\ref{fig:T90}) is very similar to those seen for SGRs
1806$-$20 and 1900+14:  161.8~ms and
93.9~ms, respectively.  \citet{gkw+01} suggested that the
difference between these values for the two SGRs is a result
of a different intrinsic physical property of the sources, such as the
strength of magnetic field, or the size of the active region.  Given
the generally softer persistent emission spectra of AXPs compared to
SGRs, as well as the less frequent outbursts of the AXPs, it is
reasonable to suspect that the two source classes differ also by some
physical property; age \citep{kds+98,gsg01}, magnetic field
\citep{gkw02,kgw+03}  and progenitor mass \citep{gsg01} have been proposed.  
The similarity of the burst
durations of all three sources implies, however,
that the physical property resulting
in different mean burst durations must be different from that
which results in different average spectra and outburst frequency.

The standard deviation of the $\tnin$ distribution for
the \tfn\ bursts is much larger than is seen for the SGR bursts.  For
\tfn, the 1$\sigma$ range is from $\sim$14~ms to $\sim$684~ms or 1.7
magnitudes. For SGRs 1806$-$20 and 1900+14, the corresponding range in 
durations is 0.68 and 0.70 magnitudes. The lower bound on the \tfn\ 
distribution may be artificially lower due to the shorter time scales
searched in this work as compared to \citet{gkw+01} who searched for SGR
bursts on the 0.125~s time scale. 
However, such a wide range of durations is seen even when faint bursts are 
omitted from the $\tnin$ distribution of \tfn.
\citet{gkw+01} argued that if the ``trapped fireball'' model, which
describes the giant SGR bursts well, also applies to the fainter
bursts, then the narrowness of the $\tnin$ distribution compared with
the wide range of fluences demands a planar fireball geometry.  This is
because the duration of the burst is limited by the rate of cooling
through the radiative fireball surface layer.  For \tfn, the $\tnin$
range is larger than the fluence range, indicating that if the fireball
model applies, a planar fireball geometry is not supported.

The distribution of burst fluences for \tfn\ is remarkably similar to those
seen in SGRs.  For the \tfn\ bursts, we find a fluence distribution $\mathrm{d}N/\mathrm{d}F \propto F^{-1.7 \pm 0.1}$ (Fig.~\ref{fig:fluence distribution}). 
\citet{gwk+00} showed that for the PCA, the fluence distribution for
SGR~1806$-$20 is well described by a power law of index $-1.43 
\pm 0.06$, while
at higher burst energies, the index steepens to $-1.7$.  For SGR~1900+14,
\citet{gwk+99} found an index of $-1.66^{+0.13}_{-0.12}$ extending over the
full range of burst fluences.  The good agreement of the fluence distribution
indices shows that for a given outburst intensity (i.e.\ the normalization of
the fluence distribution), the average burst energy 
is the same for \tfn\ as it is for these two SGRs.  
The difference between the SGR outbursts that are
routinely detected by IPN detectors and this outburst from \tfn\ which was not
detected by the IPN is the SGR outbursts have shown higher outburst
intensities.  Since we know that the SGRs spend most of their time in
quiescence when the fluence distribution normalization is zero (or near zero),
the dynamic range of the outburst intensities in SGRs is larger than has been
observed thus far in \tfn.  This difference in range is intrinsically even
larger when one considers that \tfn\ is believed to be significantly closer (3
kpc) than either of these two SGRs \citep[$\sim$15 kpc,][]{vhl+00,cwd+97}.

\citet{cegy96} noted the similarity of the fluence distribution index for
SGR~1806$-$20  with that determined empirically for earthquakes
\citep{gr56a,gr56b,gr65}, and also for the distribution of earthquake energies
found in computer simulations \citep{kat86}. However, solar flares also show a
size distribution with exponents ranging from 1.53 to 1.73
\citep{cad93,lhmb93}.  Magnetars are not clearly physically 
analogous to either
system; in magnetars, magnetic stresses are
thought to result in stellar crust cracking, which is not the case for
earthquakes.  The bursts could be magnetic reconnections as in solar flares
\citep{lyu02}, however in the solar case there is no solid crust to yield,
unlike in magnetars.  The similarity of the distributions could be
explained as being a result of the phenomena of self-organized criticality
\citep{btw88}, in which a system is dynamically attracted 
(i.e. self-organized) to a critical, spatially self-similar state 
which is just barely stable to
perturbations. In other words, the burst statistics alone do not constrain
their physical origin.

It is not possible to compare peak flux distributions as none are
published for SGRs.  For the AXP, the range of 2--60~keV peak flux for the
62.5-ms time scale spans a factor of $\sim$100, ranging from
$\sim 1 \times 10^{-9}$ to $\sim 1 \times 10^{-7}$~erg~cm$^{-2}$~s$^{-1}$.
which, for a distance of 3~kpc, corresponds to luminosities of
$\sim 1 \times 10^{36}$ to $\sim 1 \times 10^{38}$~erg~s$^{-1}$.
At time scales as short as 1/2048~s we find  peak fluxes as high as  $\sim 8 \times  10^{38}$~erg~s$^{-1}$. Thus 5 bursts are above the Eddington limit on this time scale.

As in SGRs, the fluences of the \tfn\ bursts are significantly
positively correlated with $\tnin$ (Fig.~\ref{fig:fluence versus
T90}).  However there is one difference: for the AXP, the relationship
is well described by a power law of index $+0.54 \pm 0.08$, while for
SGRs 1806$-$20 and 1900+14, \citet{gkw+01} found $+1.05\pm0.16$ and
$+0.91 \pm 0.07$, respectively.  Thus the power-law index for AXPs is
half that seen in SGRs. It is important to recognize, however,  that
severe selection effects are at work here.  Specifically, as discussed
in \S\ref{sec:duration}, we are less sensitive to low-fluence bursts.
This is particularly true for bursts having long rise times, which
will tend to have long $\tnin$ values.  Thus there are severe
selection effects against finding bursts in the bottom right-hand
portion of Figure~\ref{fig:fluence versus T90}, as there are in
similar analyses for SGRs.  Therefore the above correlation should
really be seen as an upper envelope to the phase space available
to the burst.  By contrast, our sensitivity to bursts that would sit
in the upper left-hand corner of the plot is generally enhanced relative
to the populated region, indicating the absence of bursts in this
part of phase space is genuine.

The morphologies of the AXP and SGR bursts are similar, with most
being asymmetric, with faster rises than decays.  Rise
and fall time distributions for the SGRs have not been published, so we
cannot compare those parameters directly, nor the ratio of the two.
\citet{gkw+01} showed the distribution of the ratio $t_r/\tnin$ 
for SGRs 1806$-$20 and 1900+14; the same plot for \tfn\ looks
similar (Fig. \ref{fig:ratio tr T90}).

The waiting time distributions of the AXP and SGRs are very similar.
All are well described by log-normal distributions.  This is similar to
what is seen in other self-organized critical systems, such as
earthquakes \citep{nb87}.  For \tfn, we find a mean waiting time
between bursts of 47~s, and range of 10--208~s.  \citet{gwk+99} found
$\sim$49~s for SGR 1900+14, and \citet{gwk+00} found $\sim$97~s for SGR
1806$-$20, with range between $\sim$0.1 and 1000~s for both, very
similar to our results.  The absence of correlation of waiting time and
burst fluence for the AXP is similar to that seen for SGRs
\citep{gwk+99,gwk+00}, although \citet{gwk+99} report an
anticorrelation between time since the previous burst and burst
energy.  We do not see this for the AXP, nor do \cite{gwk+00} observe
it for SGR~1806$-$20.

One striking difference between the AXP and SGR bursts is in the
relationship between spectral hardness ratio and fluence.  For
SGR~1806$-$20, \citet{gkw+01} found that the more energetic bursts are
spectrally softer, regardless of burst morphology.  This was not seen
for SGR~1900+14, however.  Our analysis (see Fig.~\ref{fig:hardness
ratio versus fluence}) shows the opposite behavior to that seen
in SGR~1806$-$20, with the more energetic bursts having harder
spectra.  \citet{gkw+01} argued that the behavior seen for SGRs could
be explained either by the emitting plasma being in local thermodynamic
equilibrium, having radiative area decreasing for lower fluences, or
by the spectral intensity of the radiation field being below that of
a blackbody, hence the emitting plasma temperature $T$ remaining in a
narrow range, being higher at lower luminosities.  Which of these two
applies depends on the rate of energy injection into the magnetosphere;
the latter applies only if the luminosity is less than $\sim 10^{42}
(V^{1/3}/10$~km)~erg~s$^{-1}$ where $V$ is the injection region,
assuming a spherical geometry.  Clearly neither can apply for the AXP.
\citet{gkw+01} imply that blackbody emission from a constant radius
predicts the relationship between hardness and fluence that we find for
the AXP.  However for the AXP, naively taking Figure~\ref{fig:fluence
versus T90} at face value, $F \propto \tnin^{0.5}$.  Hence $L_a \propto
F^{-1}$, so blackbody emission from a constant radius predicts $T
\propto F^{-1/4}$, the opposite to what we have observed.  We note
further that the range of hardness ratios for the AXP bursts is
slightly greater than it is for the SGRs.  For \tfn, hardness
ratios (for bursts having  $10^2$--$10^3$~counts) range 
from $\sim$0.54--0.85, 
while the range is $\sim$0.82--0.95 for SGR 1806$-$20, 
and $\sim 0.63$--0.67 for SGR 1900+14 \citep{gkw+01}.
It should be noted however that  we identified  bursts 
(see \S\ref{sec:searching}) using a different 
energy range (2--20~keV) than \citet{gkw+01}, 
who used the full bandpass of the PCA. 
This would make us more sensitive to softer bursts which would 
affect the dynamic range of the hardness ratios we measured.  
Perhaps interestingly, for the SGRs, $F \propto \tnin$, so the
$L_a \equiv F/\tnin \simeq $ constant, and for constant radiative area
and blackbody emission, one expects $T \simeq $ constant, closer
to what is observed for SGRs than for AXPs.  Thus, although blackbody
emission from a constant radius (not surprisingly) does not describe
any of the data well, it does seem possible that the flatter dependence of
fluence on $\tnin$, the inverted dependence of hardness on fluence
relative to the SGRs, and the greater range of hardness in the AXP
bursts may all be related phenomena telling us something interesting
about the physical distinction between these closely related sources.

We have stated that outbursts from AXPs similar to or larger 
than  the one studied here  are less frequent than are those from SGRs. 
Of course, given that we have observed only one AXP outburst, 
and that this outburst was energetically smaller and 
fainter than observed SGR outbursts, making a meaningful comparison 
of  their  outburst rate is very difficult. 
We can estimate the rate of AXP outbursts 
of the magnitude of the 2002 June 18  event  as follows.  
We consider data from only our \xte\ PCA
monitoring program, as it provides a consistent quasi-regularly sampled
data set with a single instrument.  The monitoring program for
\tfn\ has extended over nearly 7~yr with only one such outburst
detected; even though the bursting appears to have been relatively
short-lived, the effects of a glitch of even much 
smaller size would easily have been detected throughout the data span.
We make the admittedly speculative assumption that all such outbursts are 
accompanied by comparably sized glitches.
A comparable glitch in AXP \soe\  was recently
detected in 5.4~yr of monitoring 
without evidence for radiative outburst, however the sparse
observations could have missed one \citep{kg03,dis+03}.  Two small
bursts have been seen in 6.8~yr  of timing of AXP \tfe\
\citep{gkw02}, and its timing behavior suggests that many glitches could
be occurring \citep{kgc+01}, however no other evidence for radiative
outbursts has been found.  No activity of any kind, apart from
apparently simple timing noise, has been seen in 6.5~yr of timing of
\oft\  \citep{gk02} or in 4.3~yr of timing \efo\ 
\citep{ggk+02}.  If we omit  \tfe\  whose timing behavior we do not
fully understand, we can estimate a rough AXP outburst rate of one every
11~yr, assuming that the glitch in \soe\ was indeed a similar
outburst, or one every $\sim$22~yr if not. 
SGRs, by contrast, burst much more frequently, reach higher intensities, and
persist for longer periods of time.  The monitoring of the SGRs with the \xte\ PCA has not been as regular as for the AXPs due to less optimal observing
conditions for the SGRs (lower pulsed fractions, source flux, stronger timing
noise, etc.), therefore, we cannot make a direct comparison of the outburst
recurrence rate using the PCA data.  We can, however, make a rough estimate of
the recurrence rate using results obtained with the Burst and Transient Source
Experiment (BATSE) that flew aboard the {\it Compton Gamma-Ray Observatory}. 
The advantage of using BATSE to estimate the SGR outburst rate is
its uniform and dense coverage in time due to its ``all-sky'' FOV.  The
disadvantage is that BATSE is much less sensitive to SGR bursts than is the PCA
\citep[e.g.][]{gwk+99}.  Since SGR/AXP burst energies follow a steep
power-law distribution, the outburst recurrence rate is a strong function of
detector sensitivity.  It follows that an outburst recurrence rate determined
by BATSE will then be a lower limit to the rate for the more sensitive PCA. 
Moreover, the relative distances of AXPs and SGRs must be considered when
determining intrinsic source rates for a given luminosity or total energy as
opposed to peak flux and fluence.  With these factors in mind, we now estimate
the SGR outburst recurrence rate at the BATSE sensitivity level. BATSE was
in operation for 9.1~yr from 1991 April through 2000 June.  During that
time, three of the four known SGRs entered outburst 
\citep{kfm+93, kfm+94, wkv+99, gkw+01}, some
multiple times.  Here, we define an outburst as a collection of bursts (i.e.\
more than two) where the separation between consecutive bursts never exceeds
one month.  Using the results reported in \citet{gkw+01}, the number of
SGR outbursts detected during this time interval is 14.  This yields an
outburst rate for the SGRs of once every $\sim$2.6 years.  Recall, this is a
lower limit to the rate at the PCA sensitivity level. Thus the SGRs clearly 
undergo outbursts more frequently than do AXPs.

\section{CONCLUSIONS}
\label{sec:conclusions}

The bursts we have observed for \tfn\ are clearly similar to those seen 
uniquely in SGRs.  As concluded by \citet{gkw02} and \citet{kgw+03},
AXPs and SGRs clearly share a common nature, as has been predicted by
the magnetar model.  In this paper, we have done a quantitative
analysis of the \tfn\ bursts seen on 2002 June 18, and compared our results
with those obtained for the two best-studied SGRs, 1806$-$20 and 1900+14.
We summarize our results as follows.  The bursts seen in the
2002 June 18 outburst of \tfn\ are qualitatively similar to
those seen in SGRs, and in many ways quantitatively similar.
Specifically:
\begin{itemize}
\item the mean burst durations are similar
\item the differential burst fluence spectrum is well described 
by a power law of index $-1.7$, similar to those seen in SGRs (and earthquakes
and solar flares)
\item burst fluences are positively correlated with burst durations
\item the distribution of and mean waiting times are similar
\item the burst morphologies are generally asymmetric, with rise times usually
shorter than burst durations
\end{itemize}
However, there are some interesting quantitative differences between
the properties of the AXP and SGR bursts.  These may help shed light on
the physical difference(s) between these classes.  The differences can be summarized as:
\begin{itemize}
\item there is a significant correlation of burst phase with pulsed intensity, unlike in SGRs 
\item the AXP bursts have a wider range of burst duration (though this may be partly due to different analyses procedures)
\item the correlation of burst fluence with duration is flatter for AXPs than
it is for SGRs (although when selection effects are considered, this correlation
should really be seen as an upper envelope for AXPs and SGRs)
\item the fluences for the AXP bursts are generally smaller than are in observed SGR bursts
\item the more energetic AXP bursts have the hardest spectra, whereas for SGR bursts,
they have the softest spectra
\item under reasonable assumptions, SGRs undergo outbursts much 
more frequently
than do AXPs
\end{itemize}

Given the rarity of AXP bursts coupled with the unique
information that detection of such bursts provides, observing more
outbursts is obviously desirable.  Continued monitoring is thus clearly
warranted, and \xte\ with its large area and flexible scheduling
is the obvious instrument of choice. 

\acknowledgments

We are grateful to C.~Kouveliotou, M.~Lyutikov, S.~Ransom,
M.S.E.~Roberts, D.~Smith, and C.~Thompson for useful discussions.
This work was supported in part by NSERC, NATEQ, CIAR and NASA.
This research has
made use of data obtained through the High Energy Astrophysics Science
Archive Research Center Online Service, provided by the NASA/Goddard
Space Flight Center.

\bibliographystyle{apj}
%\bibliography{journals1,modrefs,psrrefs,crossrefs}

\clearpage
\begin{figure}
\plotone{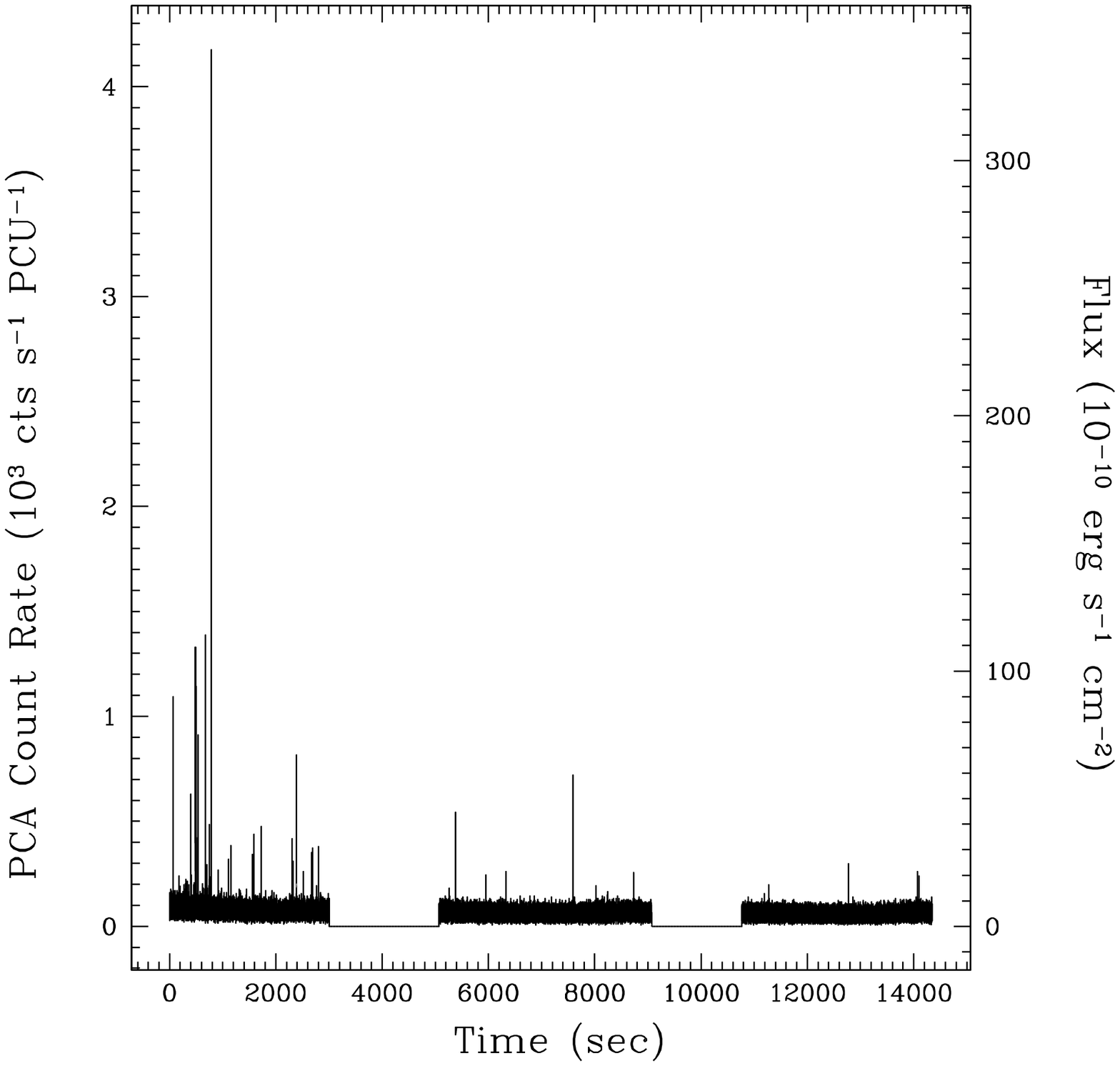}
\figcaption{2--60~keV \xte/PCA light curve for \tfn\ on 2002 June 18, at 62.5-ms resolution. The gaps are Earth occultations. \label{fig:light curve}}
\end{figure}

\clearpage
\begin{figure}
\plotone{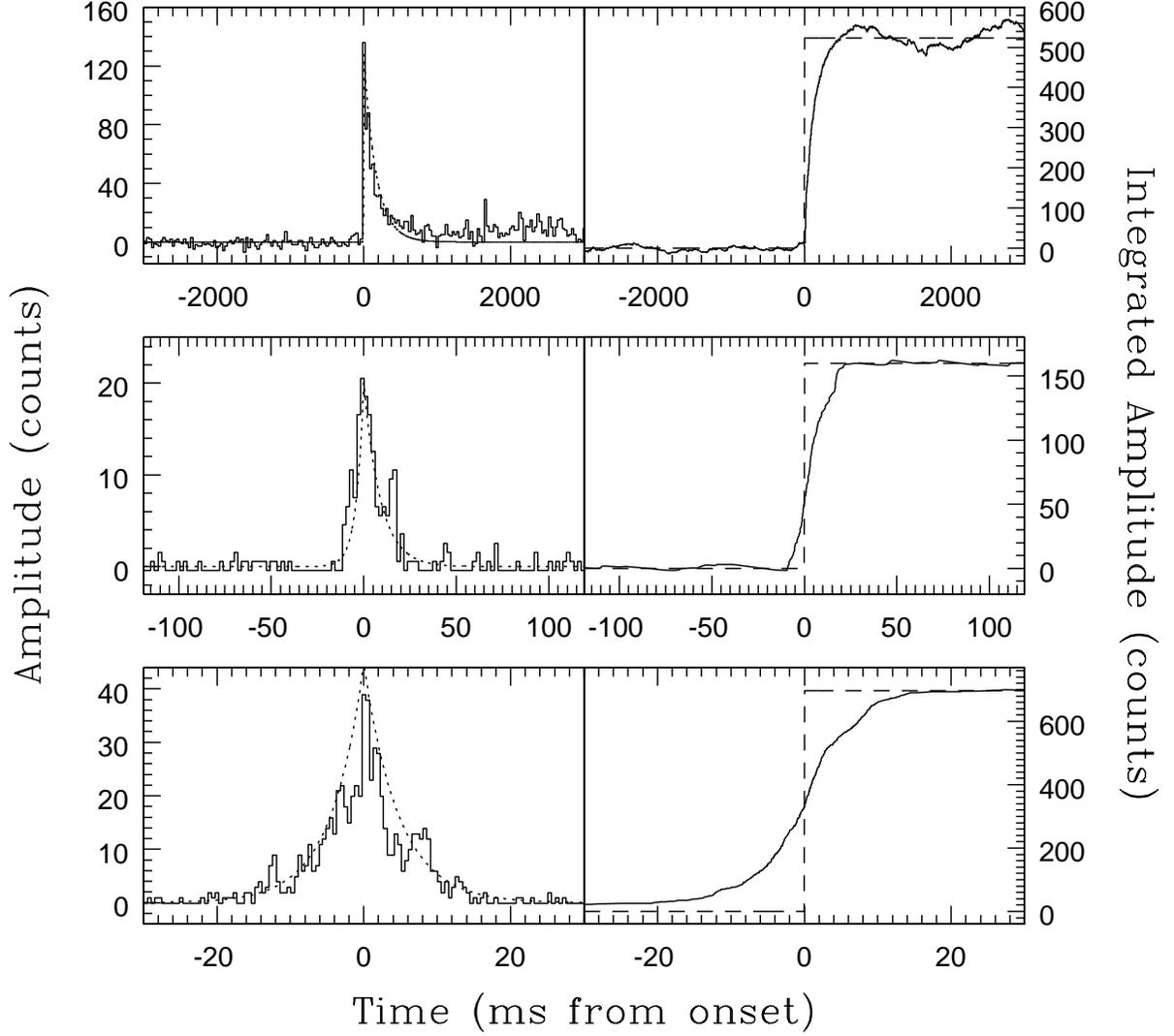} 
\figcaption{
Three different examples of bursts seen in the 2002 June 18 outburst
of \tfn.  Left: Sample background-subtracted light curves in the energy range 2--60~keV  with
1/32~s (top), 1/512~s (middle) and 1/2048~s (bottom) time resolution. 
The dotted line shows the model fit to the data in
order to measure burst rise and fall times
(see \S\ref{sec:rise fall} for details).
Right: Cumulative background-subtracted 
counts for each burst.  The vertical dotted line shows
the location of the burst peak.  The horizontal dotted line
shows the level used in determining the burst fluence.  See \S\ref{sec:duration}
for details.
\label{fig:burst profiles}}
\end{figure}

\clearpage
\begin{figure}
\plotone{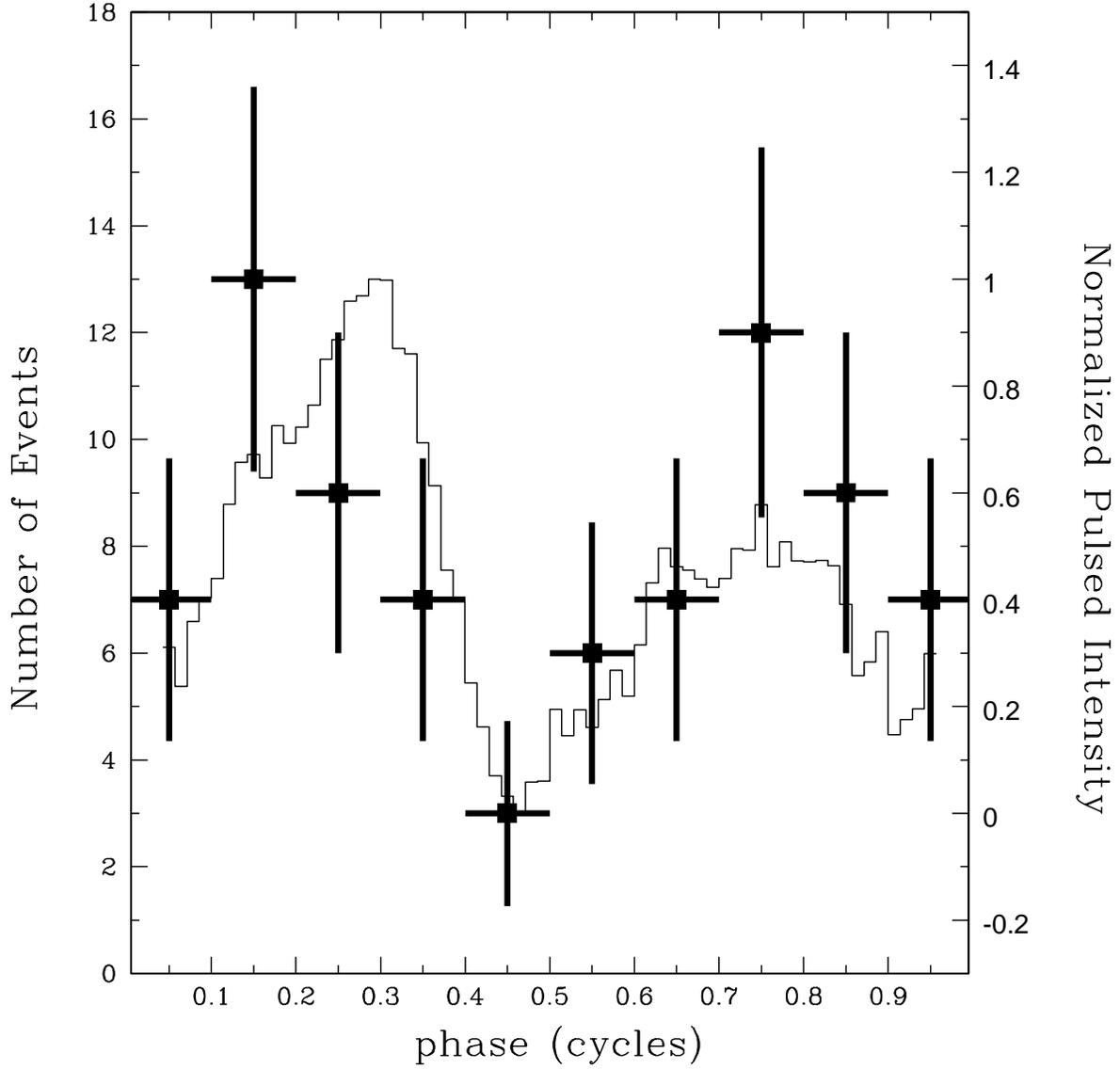} 
\figcaption{Distribution of the pulse phases of \tfn\ which correspond to the times of the burst peaks (solid points). The solid curve is the folded 2--60~keV light curve of the 2002 June 18 observation with the bursts omitted.
\label{fig:burst phases}}
\end{figure}

\clearpage
\begin{figure}
\plotone{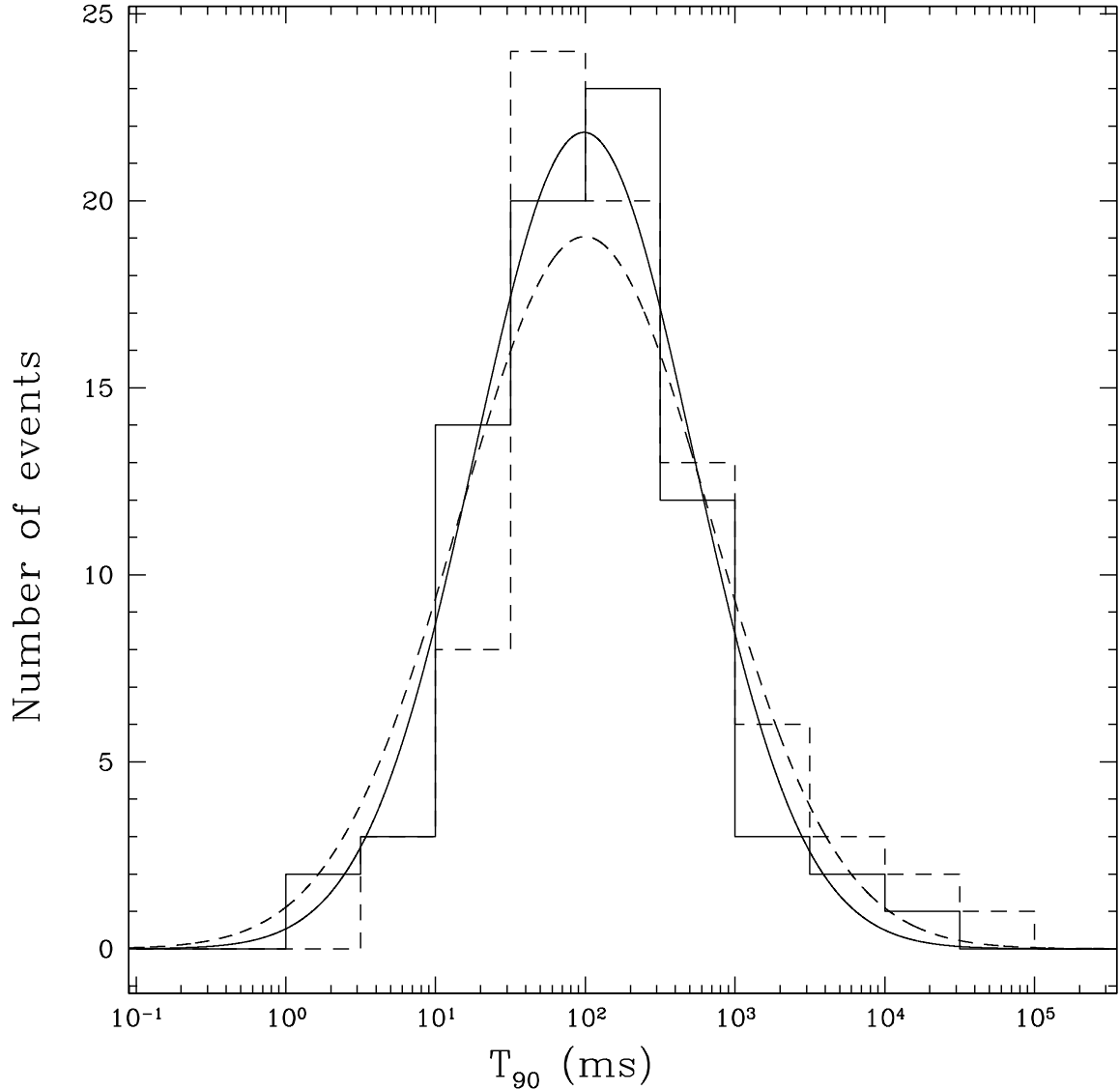} 
\figcaption{Distribution of $\tnin$ durations for the bursts observed 
from \tfn.  The solid histogram line shows the observed binned distribution 
(see \S\ref{sec:duration}), while the dashed histogram line shows the corrected 
distribution (see \S \ref{sec:t90sims}).  The solid curve 
represents the best-fit log-normal model for the observed data, as determined by
maximum-likelihood testing.  The dashed curve is the best-fit log-normal
model for the corrected data.  This fit has mean 99.31~ms and standard deviation
of a factor of 6.9.
\label{fig:T90}}
\end{figure}

\clearpage
\begin{figure}
\plotone{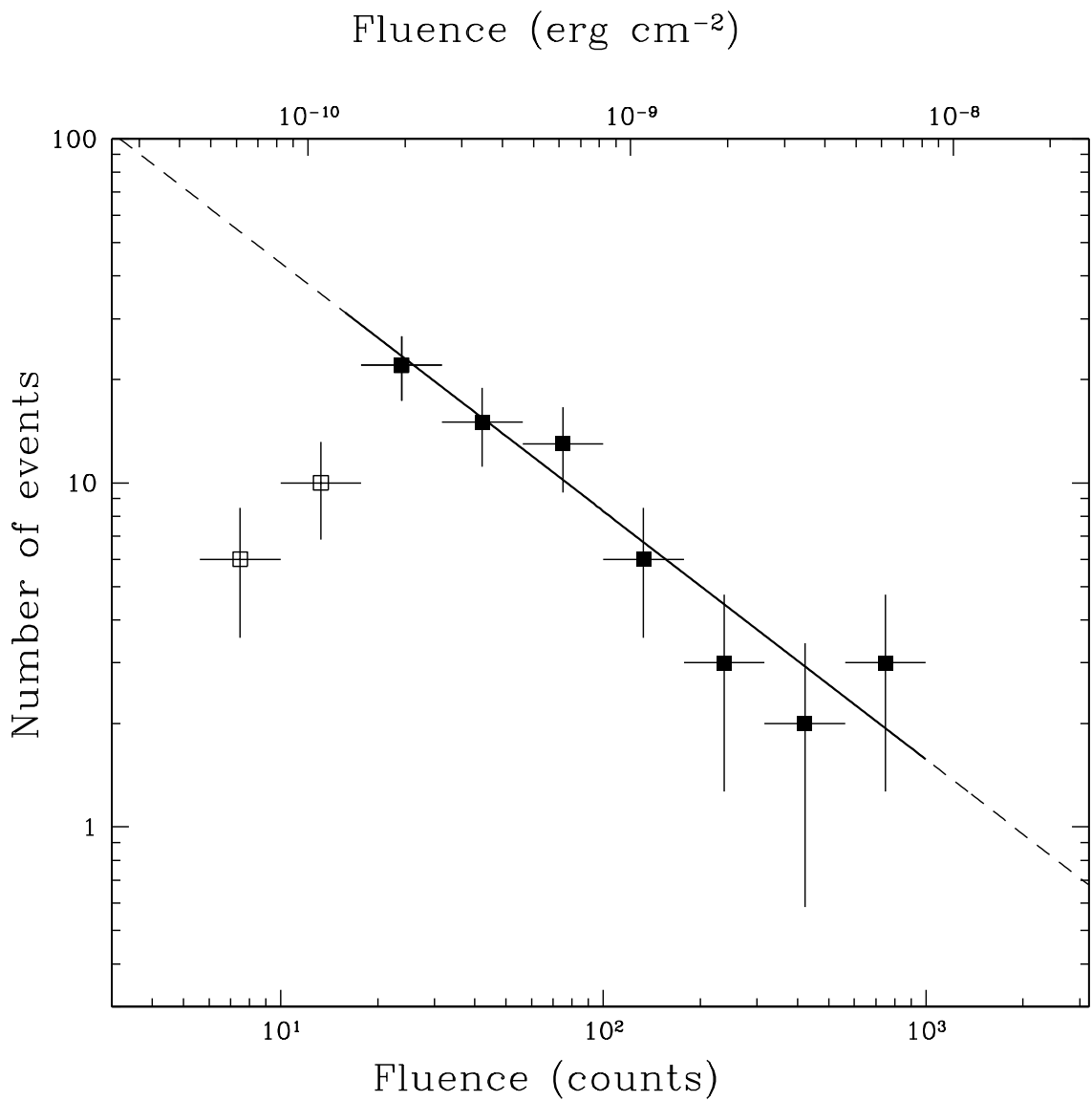} 
\figcaption{Distribution of
the 2--60~keV fluence $F$ for each burst observed from \tfn.  
Solid points represent average values of fluence in equispaced logarithmic
bins for which our observations had full sensitivity.  The open points
suffered from reduced sensitivity.  The best-fit line was determined using
the solid points only and is shown as a solid line; the dashed lines are
its extrapolation.  The slope of this line is $-0.7 \pm 0.1$, which corresponds
to d$N$/d$F \propto F^{-1.7}$.
\label{fig:fluence distribution}}
\end{figure}

\clearpage
\begin{figure}
\plotone{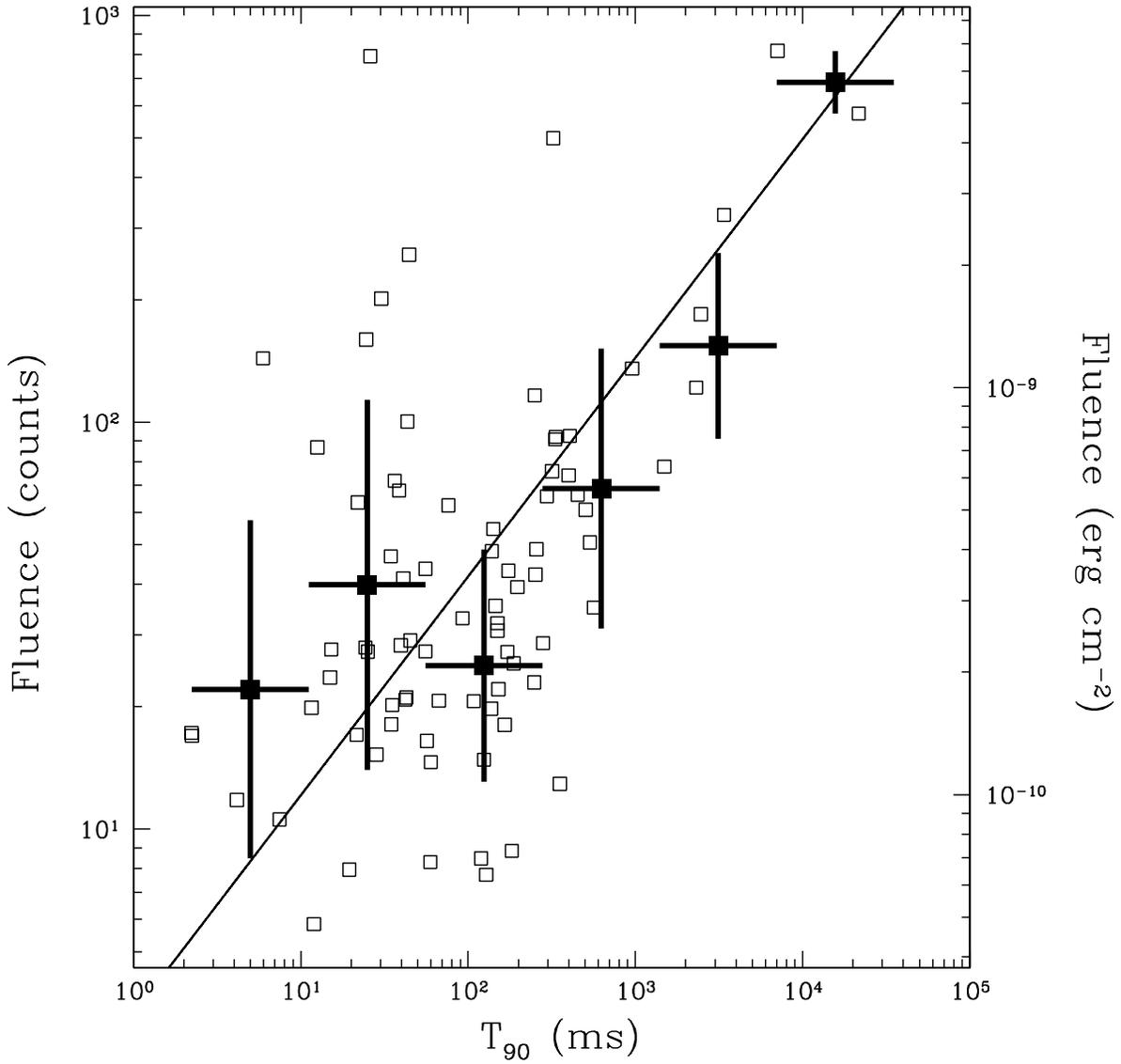} 
\figcaption{Burst 2--60~keV fluence versus $\tnin$. 
The open points represent individual bursts.  The solid points represent
binned averages.  The solid line represents the best-fit power law
for the binned averages.  
The slope of the line is $+0.54\pm0.08$.
\label{fig:fluence versus T90}}
\end{figure}

\clearpage
\begin{figure}
\plotone{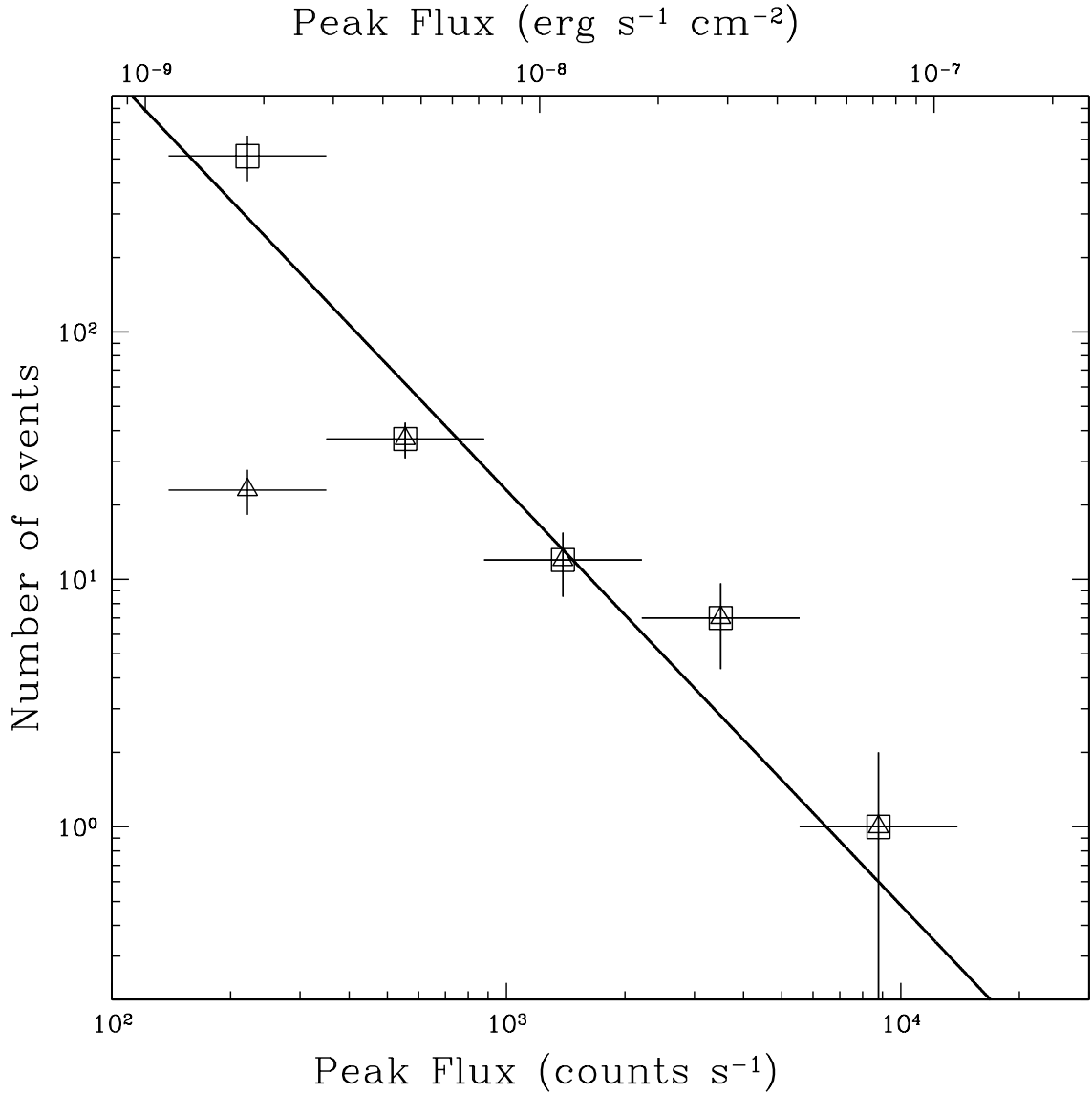} 
\figcaption{
Distribution of burst peak flux for 62.5-ms time binning.  The diamonds  
are observed averages
in equispaced logarithmic bins.  Our sensitivity is significantly reduced 
at low peak fluxes.  The corrected values, determined using
simulations described in \S~\ref{sec:flux} 
are shown by open squares. The corrected flux  bins  
were fit with a power law, shown by a line.
The slope is $-1.42 \pm 0.13$.
\label{fig:peak flux distribution}}
\end{figure}

\clearpage
\begin{figure}
\plotone{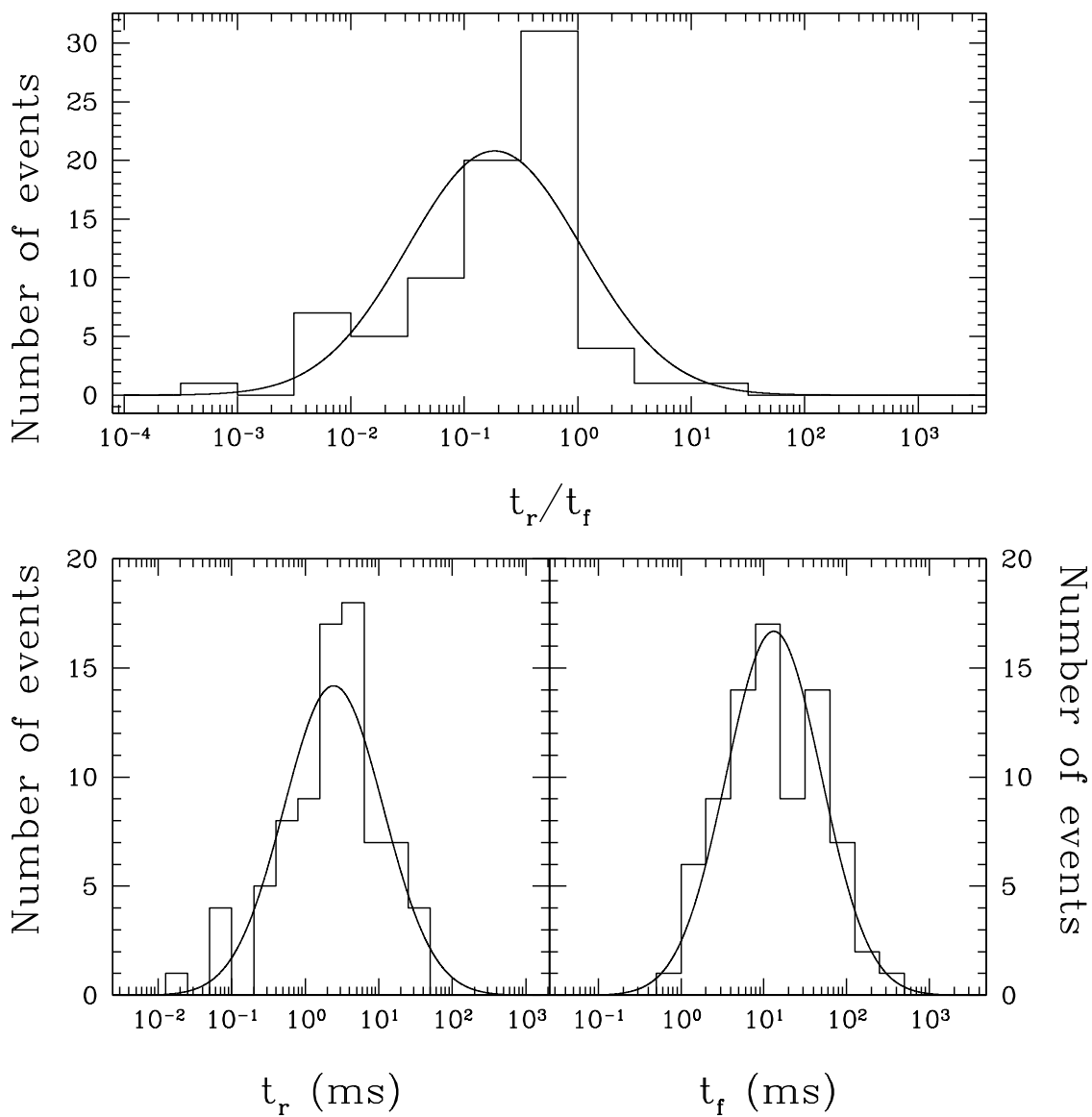} 
\figcaption{ Distribution
of burst rise ($t_r$) and fall ($t_f$) times (see \S\ref{sec:rise fall}).
Bottom left: Distribution of fall times $t_r$.  Bottom right: Distribution 
of fall times $t_f$.  Top:  Distribution of $t_r/t_f$.  In all cases, the 
solid line represents the best fit log-normal model, as determined by
maximum-likelihood testing.
\label{fig:rise fall}}
\end{figure}

\clearpage
\begin{figure}
\plotone{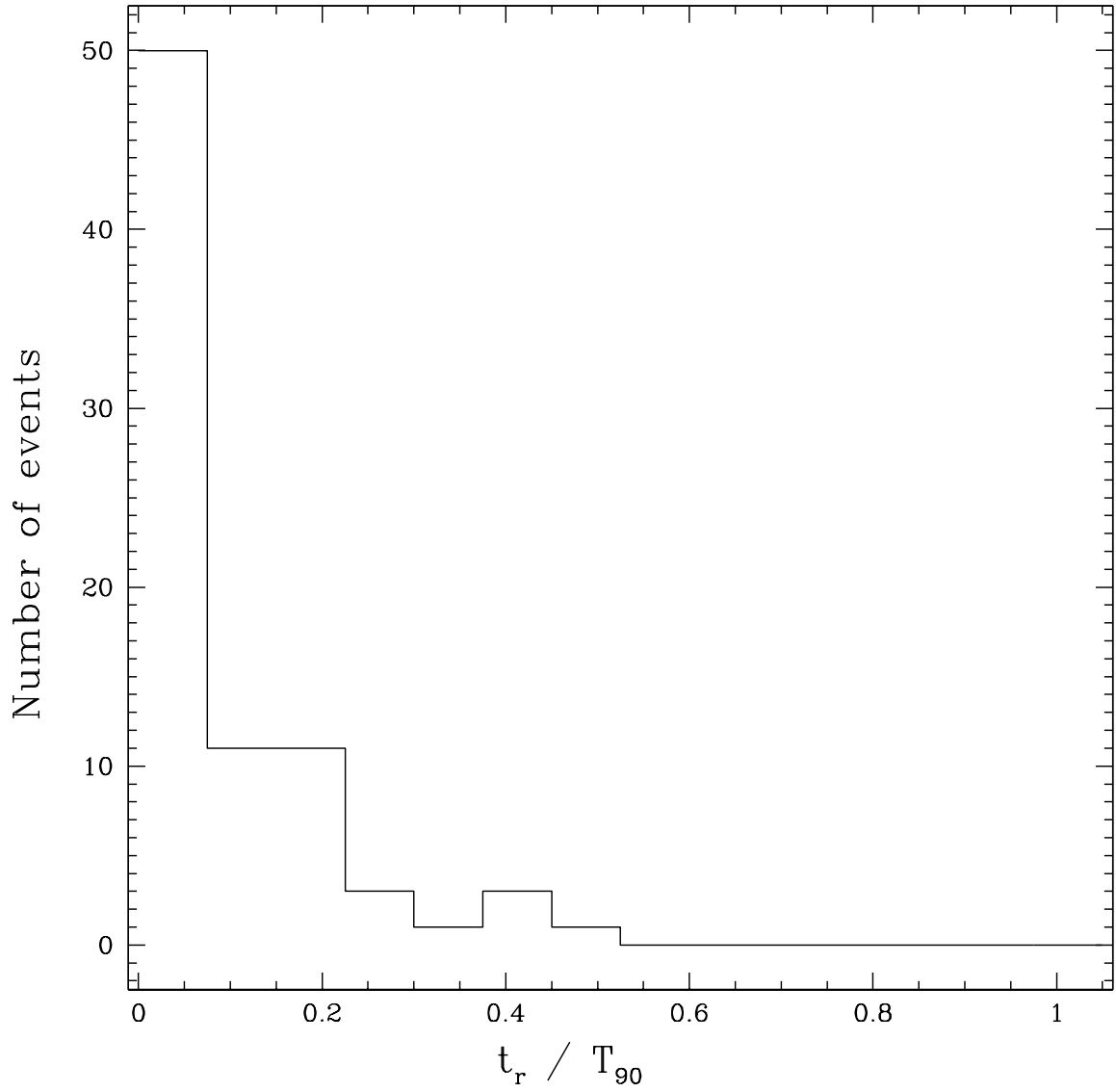} 
\figcaption{Distribution of the ratio of burst rise time $t_r$ to duration $\tnin$.
\label{fig:ratio tr T90}}
\end{figure}

\clearpage
\begin{figure}
\plotone{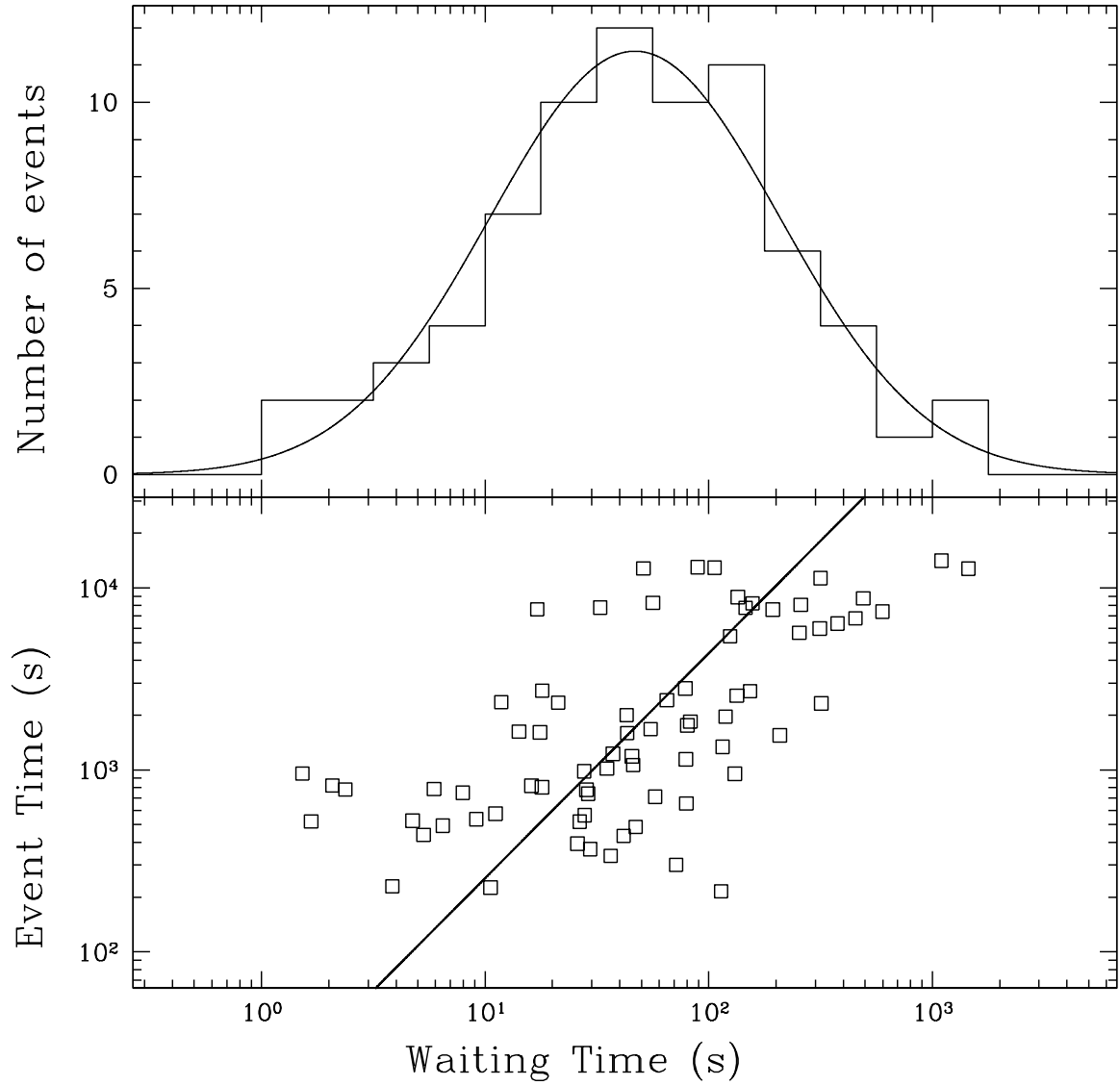} 
\figcaption{Top: Distribution
of the waiting time between successive bursts.  The solid line
represents the best fit log-normal model, as determined by
maximum-likelihood testing.   The mean is 46.8~s, and standard deviation of a factor 4.4.
Bottom: Waiting time as a function of event time. The line represents the best-fit power law model.
\label{fig:waiting time distribution}}
\end{figure}

\clearpage
\begin{figure}
\plotone{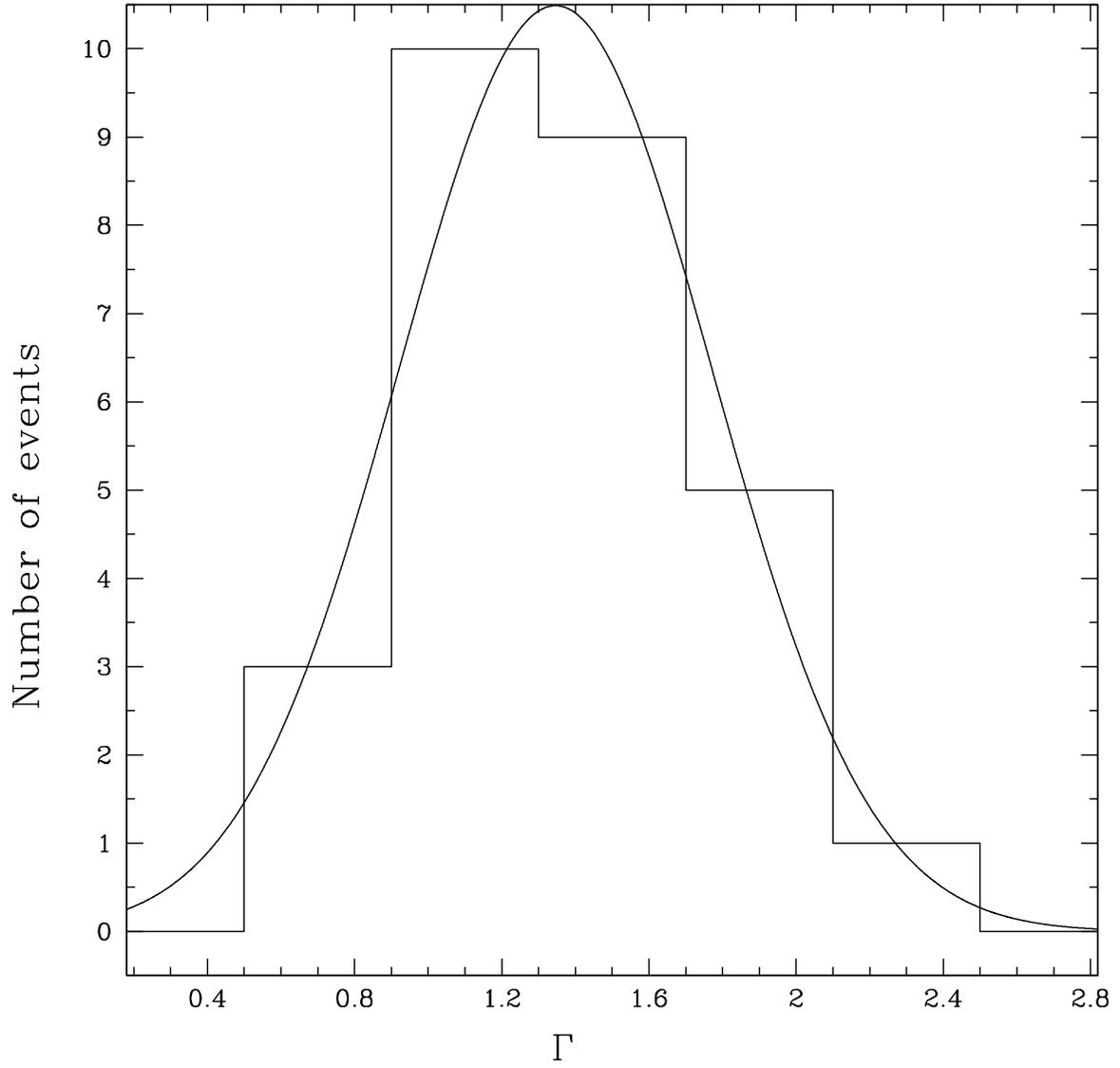}
\figcaption{Distribution of spectral indices ($\Gamma$) for the 28 most fluent bursts. See \S\ref{sec:ind burst spectra} for details.
The curve is the best-fit gaussian
model.  This fit has mean 1.35 and standard deviation 0.43. 
\label{fig:gamma distribution}}
\end{figure}

\clearpage
\begin{figure}
\plotone{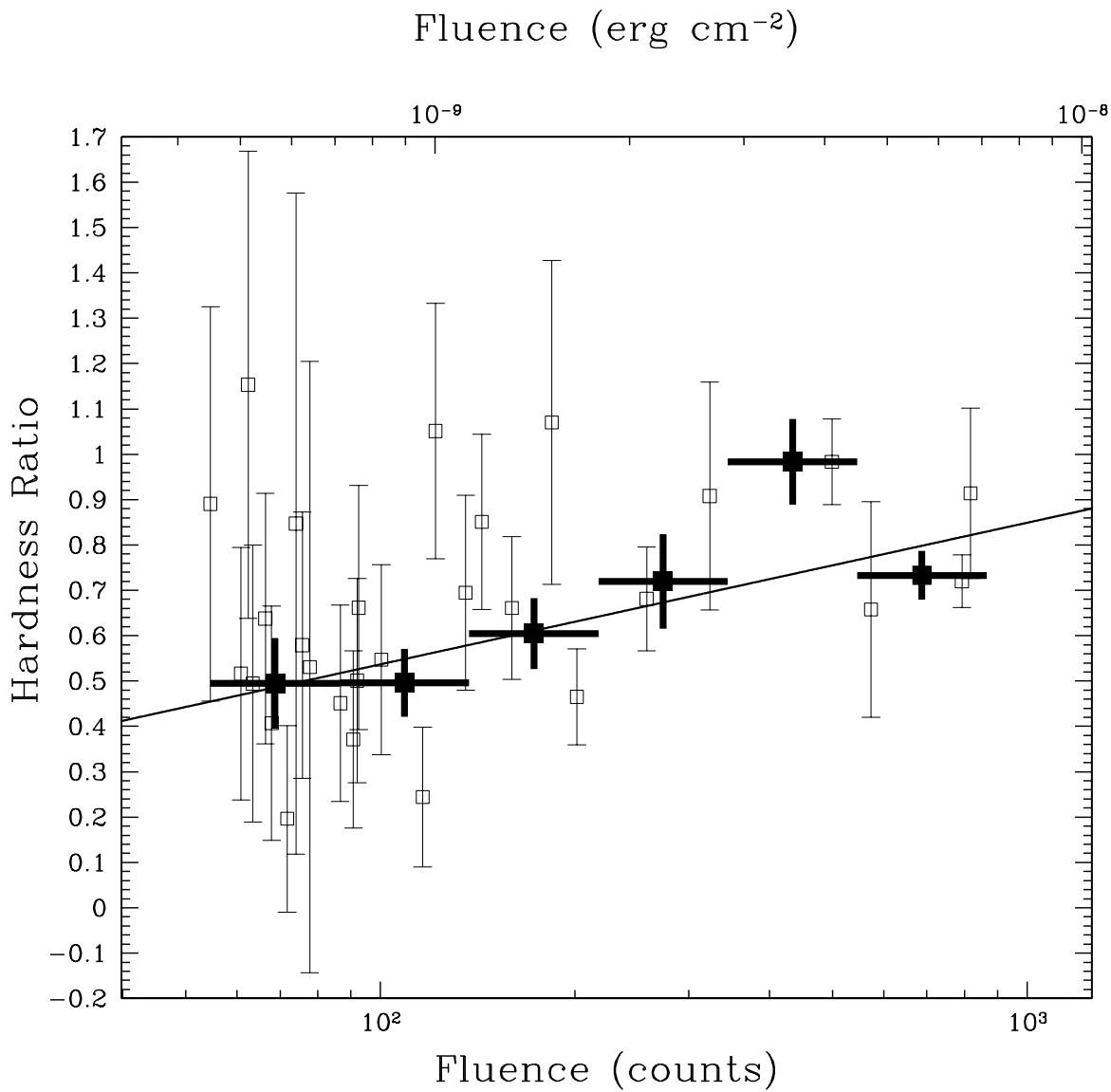} 
\figcaption{Hardness ratio ($H$) versus fluence ($F$).  Hardness ratio is defined as
the ratio of the number of PCA counts in the 10--60~keV band to that in
the 2--10~keV band.  The open points are hardness ratio measurements
for individual bursts. The solid points are weighted averages of
hardness ratios for bursts in equispaced logarithmic fluence bins. The line represents the best-fit logarithmic function for the weighted averages, $H = 0.31\times \log F - 0.09 $.
\label{fig:hardness ratio versus fluence}}
\end{figure}

\end{document}